\documentclass[sigconf,natbib=true,anonymous=false]{acmart}

\usepackage{graphicx}
\usepackage{amsmath}
\usepackage{booktabs}
\usepackage{algorithm}
\usepackage{algorithmic}
\usepackage{amsfonts}
\usepackage{multirow}
\usepackage{makecell}
\usepackage{subfigure}
\usepackage{color}
\usepackage{bm}
\usepackage{epstopdf}
\usepackage{url}
\usepackage[cal=cm]{mathalfa}
\usepackage{balance}
\usepackage{threeparttable}
\usepackage{wrapfig}

\usepackage{enumitem}
\setlist[itemize]{leftmargin=*}

\makeatletter
\newcommand\notsotiny{\@setfontsize\notsotiny\@vipt\@viipt}
\makeatother

\AtBeginDocument{%
  \providecommand\BibTeX{{%
    \normalfont B\kern-0.5em{\scshape i\kern-0.25em b}\kern-0.8em\TeX}}}

\author{Xiaodong Li}
\affiliation{
  \institution{Institute of Information Engineering, Chinese Academy of Sciences. \\School of Cyber Security, UCAS$^\dag$}
  \country{lixiaodong@iie.ac.cn}
 }

\author{Jiawei Sheng}
\affiliation{
  \institution{Institute of Information Engineering, Chinese Academy of Sciences}
  \country{shengjiawei@iie.ac.cn}
 }
 \authornote{Corresponding author.\\ $^\dag$University of Chinese Academy of Sciences.}

\author{Jiangxia Cao}
\affiliation{
  \institution{Kuaishou Inc}
  \country{caojiangxia@kuaishou.com}
 }

\author{Wenyuan Zhang}
\affiliation{
  \institution{Institute of Information Engineering, Chinese Academy of Sciences. \\ School of Cyber Security, UCAS}
  \country{zhangwenyuan@iie.ac.cn}
 }

\author{Quangang Li}
\affiliation{
  \institution{Institute of Information Engineering, Chinese Academy of Sciences}
  \country{liquangang@iie.ac.cn}
}

\author{Tingwen Liu}
\affiliation{
  \institution{Institute of Information Engineering, Chinese Academy of Sciences. \\ School of Cyber Security, UCAS}
  \country{liutingwen@iie.ac.cn}
}
\authornotemark[1]

\copyrightyear{2024}
\acmYear{2024}
\setcopyright{rightsretained}
\acmConference[WSDM '24]{Proceedings of the 17th ACM International Conference on Web Search and Data Mining}{March 4--8, 2024}{Merida, Mexico}
\acmBooktitle{Proceedings of the 17th ACM International Conference on Web Search and Data Mining (WSDM '24), March 4--8, 2024, Merida, Mexico}\acmDOI{10.1145/3616855.3635794}
\acmISBN{979-8-4007-0371-3/24/03}

\begin{document}
\title{CDRNP: Cross-Domain Recommendation to Cold-Start Users \\ via Neural Process}

\renewcommand{\shorttitle}{CDRNP}
\renewcommand{\shortauthors}{Xiaodong Li et al.}

\begin{abstract}
Cross-domain recommendation (CDR) has been proven as a promising way to tackle the user cold-start problem, which aims to make recommendations for users in the target domain by transferring the user preference derived from the source domain.
Traditional CDR studies follow the embedding and mapping (EMCDR) paradigm, which transfers user representations from the source to target domain by learning a user-shared mapping function, neglecting the user-specific preference.
Recent CDR studies attempt to learn user-specific mapping functions in meta-learning paradigm, which regards each user's CDR as an individual task, but neglects the preference correlations among users, limiting the beneficial information for user representations.
Moreover, both of the paradigms neglect the explicit user-item interactions from both domains during the mapping process.
To address the above issues, this paper proposes a novel CDR framework with neural process (NP), termed as CDRNP.
Particularly, it develops the meta-learning paradigm to leverage user-specific preference, and further introduces a stochastic process by NP to capture the preference correlations among the overlapping and cold-start users, thus generating more powerful mapping functions by mapping the user-specific preference and common preference correlations to a predictive probability distribution.
In addition, we also introduce a preference remainer to enhance the common preference from the overlapping users, and finally devises an adaptive conditional decoder with preference modulation to make prediction for cold-start users with items in the target domain.
Experimental results demonstrate that CDRNP outperforms previous SOTA methods in three real-world CDR scenarios.

\end{abstract}



\maketitle

\section{Introduction}
\begin{figure*}[t!]
	\begin{center}
		\includegraphics[width=13.5cm]{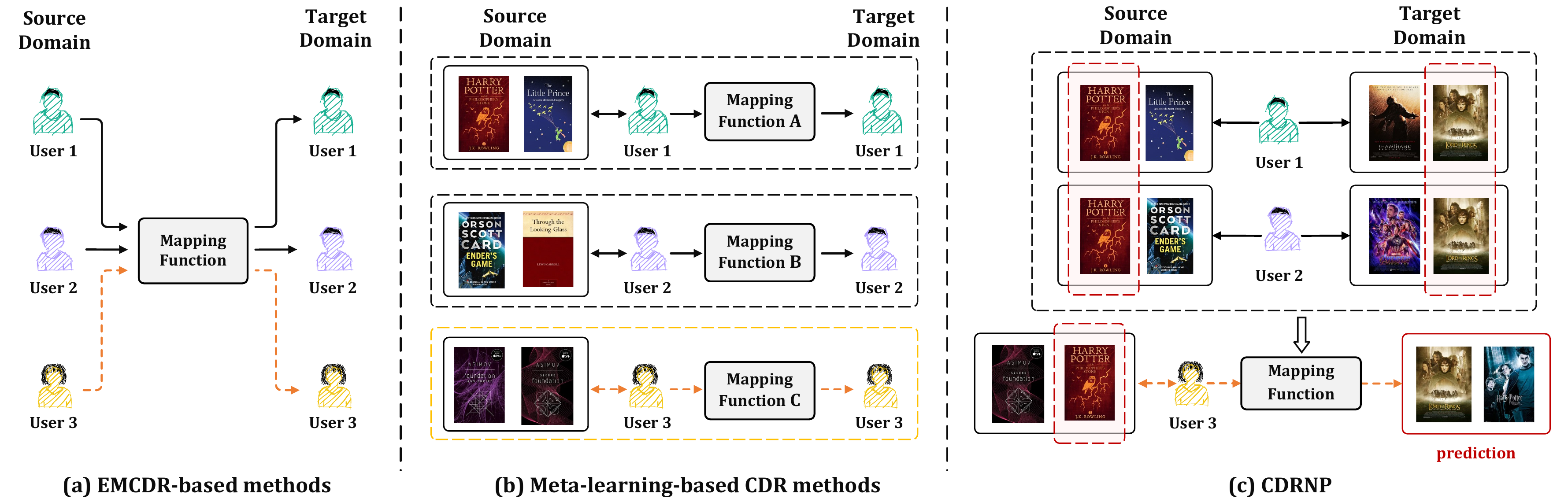}
		\caption{Users 1 and 2 are overlapping users, while User 3 is a cold-start user. (a) The framework of EMCDR-based methods by learning a user-shared mapping function. (b) The framework of meta-learning-based CDR methods by learning user-specific mapping functions. (c) The framework of our CDRNP by capturing preference correlations among users.}
		\label{fig:motivation}
	\end{center}
\end{figure*}

Recommendation systems (RS) have been widely deployed in real-world applications, such as Amazon (E-commerce) and YouTube (online video). 
As a promising method, collaborative filtering (CF) has been successfully applied to RS, ranging from early matrix factorization methods~\cite{BPR,cmf} to the recent deep neural network methods~\cite{ncf,ngcf}. 
However, CF-based methods usually suffer from the long-standing challenge of \textit{user cold-start problem}~\cite{tmcdr}, where the new emerging user has few \textit{user-item interactions} that can hardly learn effective representations to make recommendation. 
To address this problem, \textbf{\textit{cross-domain recommendation (CDR)}}~\cite{ddtcdr,cdrbtgcf} has been proposed recently, which explores the cold-start users by utilizing the rich user-item interactions observed from an informative \textit{source domain}, so as to improve the cold-start users' recommendation in the \textit{target domain}.

A common idea of CDR methods is to train the model with the \textit{overlapping users} appearing in both the source and target domain, and then make recommendation for the \textit{cold-start users} that only have interactions in the source domain.
To achieve the above idea, traditional CDR methods~\cite{emcdr,sscdr,tmcdr,tsvcdr,cdrva} follow the Embedding and Mapping (EMCDR) paradigm (as shown in Fig.~\ref{fig:motivation}(a)).
This paradigm aims to learn a user-shared mapping function. For implementation, EMCDR first generates user/item representations for each domain separately, and then trains a mapping function to align the overlapping users representations between the source and target domain. Afterward, given a cold-start user representation from the source domain, EMCDR uses the mapping function to predict the user representations in the target domain.
However, such a paradigm learns a unified mapping function shared by all users, neglecting the different user-specific preferences, thus leading to inflexible representation transferring for different users.

Recent studies attempt meta-learning~\cite{mamlfa,metanet} to further capture user-specific preference in CDR (as shown in Fig.~\ref{fig:motivation}(b)).
They mostly regard each user's CDR as an individual task, and train the model to achieve these user-specific tasks with the users in the support set, thus can make recommendation for cold-start users in the query set.
Among them, PTUPCDR~\cite{ptupcdr} proposes a meta network with a task-oriented optimization procedure, and generates cold-start user representation by personalized mapping functions in prediction.
However, this kind of methods usually regard each user's CDR as an independent task, neglecting the preference correlations among users, where the similar users may have similar preferences which can provide beneficial information for training the mapping functions. 
Moreover, both of the EMCDR-based and the meta-learning-based methods mainly concentrate on the transferring of the learned user representations, neglecting the explicit user-item interactions from both domains during the mapping process, thus leading to sub-optimal user preference modeling.

To address the above issues, we propose a novel meta-learning framework for \underline{CDR} based on \underline{N}eural \underline{P}rocess~\cite{np,cnp}, termed as CDRNP (as shown in Fig.~\ref{fig:motivation}(c)). 
Specifically, we build our model with Neural Process (NP) principle, which formulates the CDR task of all users within the same stochastic process, naturally capturing the preference correlations among the overlapping and cold-start users.
Besides, the model also follows the meta-learning paradigm, which not only preserves user-specific information for each user, but also utilizes explicit user-item interactions from both domains beyond the sole user representation transferring. Such a design enables CDRNP to learn more representative user-specific mapping functions and achieve bi-directional recommendation with same observed data. 
For implementation of the CDRNP framework, we first propose a \textit{characteristic embedding layer} which leverages each user's interactions observed in the source domain to generate its user-specific preference representation, and then combine it with the candidate item in the target domain, constituting the complete rating-observed information as input for each candidate item rating.
Then, we propose a \textit{preference aggregator} to approximate CDR for each user as a particular instance of a stochastic process, which captures common preference correlations among the cold-start users and overlapping users.  
Additionally, we devise a \textit{preference remainer} to enhance the common preference representations from the overlapping users.
Finally, we devise an \textit{adaptive conditional decoder} to modulate both preference representations to make better rating prediction for the cold-start users in the target domain.

In summary, the contributions of this paper are as follows:
\begin{itemize}
    \item To our best knowledge, we introduce a fresh perspective to solve the cold-start problem in CDR with NP principle.
    \item We propose a novel meta-learning CDR method with NP, termed as CDRNP, which preserves the capability of user-specific mapping function by meta-learning paradigm, while further captures preference correlations among users with NP.
    \item Extensive experiments on three CDR scenarios demonstrate that CDRNP outperforms the state-of-the-art methods.
\end{itemize}

\section{Preliminary}
This section introduces the CDR problem settings and the background knowledge of the neural process principle.

\subsection{Problem Setting}

This paper considers a general CDR scenario:
given the observed user-item interactions of both domains, we aim to make recommendations in the target domain for cold-start users, with the help of user-item interactions in the source domain.
Formally, the data can be formulated as $\mathcal{D}^{s/t} = (\mathcal{U}^{s/t},\mathcal{V}^{s/t},\mathcal{Y}^{s/t})$, where the superscript $s/t$ denotes the source/target domain, and $\mathcal{U}$, $\mathcal{V}$ and $\mathcal{Y}\in \{0,1,2,3,4,5\}^{|\mathcal{U}|\times |\mathcal{V}|}$ denote the user set, item set and user-item rating matrix, respectively. 
Here the rating score $y(u,v)\in \mathcal{Y}$ ranges from 0 to 5, reflecting the preference tendency of user $u\in \mathcal{U}$ to choose item $v\in \mathcal{V}$.
Further, we define the users overlapped in both domains as the overlapping users $\mathcal{U}^o = \mathcal{U}^s \cap \mathcal{U}^t$, then the cold-start users from the source domain can be denoted as $\mathcal{U}^{s\setminus o} = \mathcal{U}^{s}\setminus \mathcal{U}^o$.
Therefore, for each cold-start user $u^{s\setminus o}\in \mathcal{U}^{s\setminus o}$, given its interactions from the source domain $\mathcal{V}^{s}(u^{s\setminus o})$, the goal is to predict the rating $y(u^{s\setminus o}, v^{t})$ for each candidate $v^{t} \in \mathcal{V}^{t}$.

To achieve the above goal, we develop our model with meta-learning settings~\cite{melu,mlcri,mlcdr,mlhin,mamo} for users:
For each user $u_j\in \mathcal{U}$, to predict rating $y(u_j,v_k^{t})$ in the target domain, we briefly summarize the rating-observed information as $x_{jk} = (u_j, \mathcal{V}^{s}(u_j), v_k^t)$.
Here we use $x_{m}\triangleq x_{jk}$ and $y^{t}_{m} \triangleq y(u_j,v_k^{t})$ for brevity.
Based upon, we define the concept of task $\omega_i$ in meta-learning, where each $\omega_i = \{(x_{m}, y^{t}_{m})|(x_{m}, y^{t}_{m}) \in \{\mathcal{C}_{i}\cup \mathcal{Q}_{i}\} \}$ batched with multiple users. 
Here, we divide $\omega_i$ into the support set $\mathcal{C}_i$ and the query set $\mathcal{Q}_i$, such that $\omega_i = \mathcal{C}_i\cup \mathcal{Q}_i$.

Note that we train our model on the tasks $\omega_i\in \Omega_{meta-training}$ consisting of the overlapping users, i.e., we use $(x_{m}^{o}, y^{t}_{m})$ pairs with $u_j^o\in \mathcal{U}^o$ in both $\mathcal{C}_i$ and $\mathcal{Q}_i$, so as to learn the common knowledge of achieving CDR with these tasks.
In the testing phase, our model predicts the interactions of the cold-start users in target domain via new tasks $\omega_i \in \Omega_{meta-testing}$, where we use $(x_{m}^{o}, y^{t}_{m})$ pairs with $u_j^o\in \mathcal{U}^o$ in $\mathcal{C}_i$ and we predict $(x_{m}^{s\setminus o}, y^{t}_{m})$ pairs with $u_j^{s\setminus o}\in \mathcal{U}^{s\setminus o}$ in $\mathcal{Q}_i$.
In such a way, the model learns common knowledge of achieving CDR in the training phase, thus can make recommendation for the cold-start users to achieve CDR in the testing phase.

\subsection{Neural Process Principle}
To achieve CDR, we develop our model based on Neural Process (NP)~\cite{np}.
Generally, NP is a kind of neural latent variable models, which approximates stochastic processes parameterized by powerful neural networks.
In this paper, we assume each task $\omega_i = \{(x_{m}, y^{t}_{m})|(x_{m}, y^{t}_{m})\in \{\mathcal{C}_{i}\cup \mathcal{Q}_{i}\} \}$ associated with an instantiation of the stochastic process $f_i$, and each pair $(x_{m}, y^{t}_{m})$ can be seen as a sample point drawn from $f_i$.
Therefore, the corresponding probability distribution $\rho$ of $f_i$ can be defined as follows:
\begin{equation}
\small
        \mathbf{\rho}_{x_{1:M_i}}(y_{1:M_i}^t) = \int p(f_i)p(y_{1:M_i}^t|f_i,x_{1:M_i})df_i,
        \label{sec_math_eq:graph}
\end{equation}
where $x_{1:M_i}$ and $y_{1:M_i}^t$ denote the all $x_{m}$ and $y^{t}_{m}$ decoupled from $\omega_i$, respectively, and we use $M_i=|\omega_i|$ for simplicity.
Following  NP, we approximate the stochastic process $f_i$ with a low-dimensional random vector $\bm{z}_i$ generated by non-linear neural networks.
Then, with $i.i.d.$ condition on tasks, we rewrite the generation process as:
\begin{equation}
\small
        \begin{split}
        p(y_{1:M_i|}^t|x_{1:M_i}) = \int p(\bm{z}_i)\prod_{m=1}^{M_i} p(y_m^t|x_m,\bm{z}_i)d\bm{z}_i,
        \end{split}
        \label{sec_math_eq:graph}
\end{equation}
Here we assume $p(\bm{z}_i)\sim \mathcal{N}(\bm{0},\bm{I})$ for simplicity in paradigmatic variational frameworks~\cite{aevb}, and then we can easily sample $\bm{z}_i$ from $p(\bm{z}_i)$ to achieve concrete function implementation.

Since the true posterior distribution is intractable, we adopt amortised variational inference \cite{nvi} to learn it. Let $q(\bm{z}_i|\omega_i)$ be the variational posterior of the latent variable $\bm{z}_i$, the corresponding evidence lower-bound (ELBO) derivation is as follows:
\begin{equation}
\small
        \begin{split}
        &\log p(y_{1:M_i}^t|x_{1:M_i}) = \log\int p(\bm{z}_i, y_{1:M_i}^t|x_{1:M_i})d\bm{z}_i,\\
        &\geqslant  E_{q(\bm{z}_i|\omega_i)}\left[\log\frac{p(\bm{z}_i,y_{1:M_i}^t|x_{1:M_i})}{q(\bm{z}_i|\omega_i)}\right],\\
        &= E_{q(\bm{z}_i|\omega_i)}\left[\sum_{m=1}^{M_i} {\log p(y_m^t|x_m,\bm{z}_i)}+\log\frac{p(\bm{z}_i)}{q(\bm{z}_i|\omega_i)}\right],
        \end{split}
        \label{eq:np_meta_ori}
\end{equation}
Considering that each task $\omega_i$ contains a support set $\mathcal{C}_i$ and a query set $\mathcal{Q}_i$, we would like to make predictions for $\mathcal{Q}_i$ given $\mathcal{C}_i$.
Therefore, we rewrite Eq.~(\ref{eq:np_meta_ori}) as follows:
{\small
\begin{align}
    &\log p(y_{{1}:|\mathcal{Q}_i|}^t|x_{{1}:|\mathcal{Q}_i|},\mathcal{C}_i)\label{eq:np_log_meta}\\
    &\!\geqslant\! E_{q(\bm{z}_i|\mathcal{Q}_i)}\left[\sum_{m=1}^{|\mathcal{Q}_i|}{\log p(y_m^t|x_m,\bm{z}_i)}+\log\frac{p(\bm{z}_i|\mathcal{C}_i)}{q(\bm{z}_i|\mathcal{Q}_i)}\right], \label{eq:np_lowerbound_meta}\\
    &\!=\!-\![\!\underbrace{-E_{q(z_i|\mathcal{Q}_i)}\!\log p(y_{1:|\mathcal{Q}_i|}^t|x_{1:|\mathcal{Q}_i|},z_i)}_{\text{desired log-likelihood}}+\underbrace{\mathrm{KL}(q(z_i|\mathcal{Q}_i)||q(z_i|\mathcal{C}_i))}_{\text{KL divergence}}]\label{eq:np_loss}
\end{align}}
where Eq.~(\ref{eq:np_lowerbound_meta}) is the lower bound of Eq.~(\ref{eq:np_log_meta}) with derivation like Eq.~(\ref{eq:np_meta_ori}). 
In practice, $p(\bm{z}_i|\mathcal{C}_i)$ in Eq.~(\ref{eq:np_lowerbound_meta}) is intractable, thus we use variational prior $q(\bm{z}_i|\mathcal{C}_i)$ to approximate $p(\bm{z}_i|\mathcal{C}_i)$.
Such an approximation can be regarded as a regularization term of KL divergence to ensure the $consistency$ condition of stochastic process~\cite{stochastic}. 
Thereby, we summarize Eq.~(\ref{eq:np_lowerbound_meta}) into Eq.~(\ref{eq:np_loss}), which can be intuitively explained with two terms: the first term aims to achieve the desired CDR task, and the second term aims to ensure all tasks in the query and support set following the same stochastic process.

\begin{figure}[t!]
	\begin{center}
            \includegraphics[width=5.5cm]{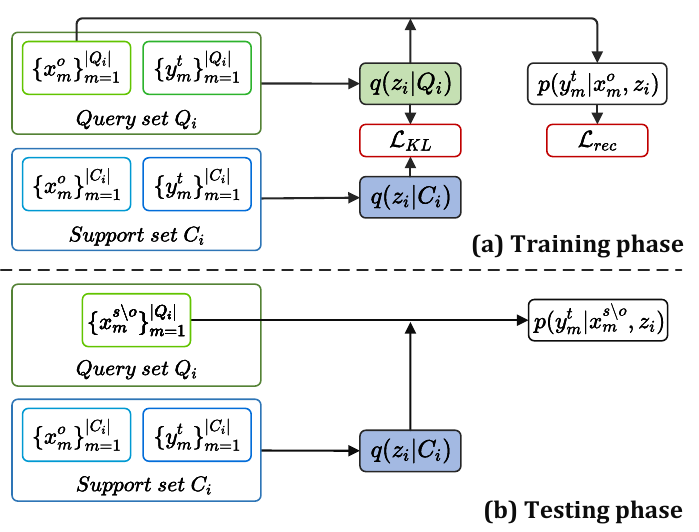}
		\caption{An illustration of our model CDRNP with NP in the training and testing phase.}
		\label{fig:train_test}
	\end{center}
\end{figure}

For implementation, we illustrate the above principle from a practical perspective.
Fig.~\ref{fig:train_test} shows our model with NP in the training and testing phase.
Note that in the training phase, we use $\mathcal{L}_{KL}$ to assimilate $q(\bm{z}_i|\mathcal{Q}_i)$ and $q(\bm{z}_i|\mathcal{C}_i)$, which learns $\bm{z}_i$ general to both the support and query set in each task.
In the testing phase, we use $\bm{z}_i$ generated from the  $q(\bm{z}_i|\mathcal{C}_i)$ to substitute for $q(\bm{z}_i|\mathcal{Q}_i)$, and thus together with $\{x_m^{s\setminus o}\}^{|\mathcal{Q}_i|}_{m=1}$ in the given task, we make predictions for the users by $p(y|x,z)$.
In this way\footnote{Note that in the training phase, we leverage overlapping users $u_j^o$ in both $\mathcal{Q}_i$ and $\mathcal{C}_i$, so we can make prediction on $\mathcal{Q}_i$ by imitating $\mathcal{C}_i$.
In the testing phase, we make predictions for cold-start users $u_j^{s\setminus o}$ in $\mathcal{Q}_i$, by imitating overlapping users $u_j^o$ in $\mathcal{C}_i$.
}, we encourage the tasks of users following the same stochastic process, thus naturally captures the preference correlations among overlapping and cold-start users.

\begin{figure*}[!t]
	\begin{center}
            \includegraphics[width=13.5cm]{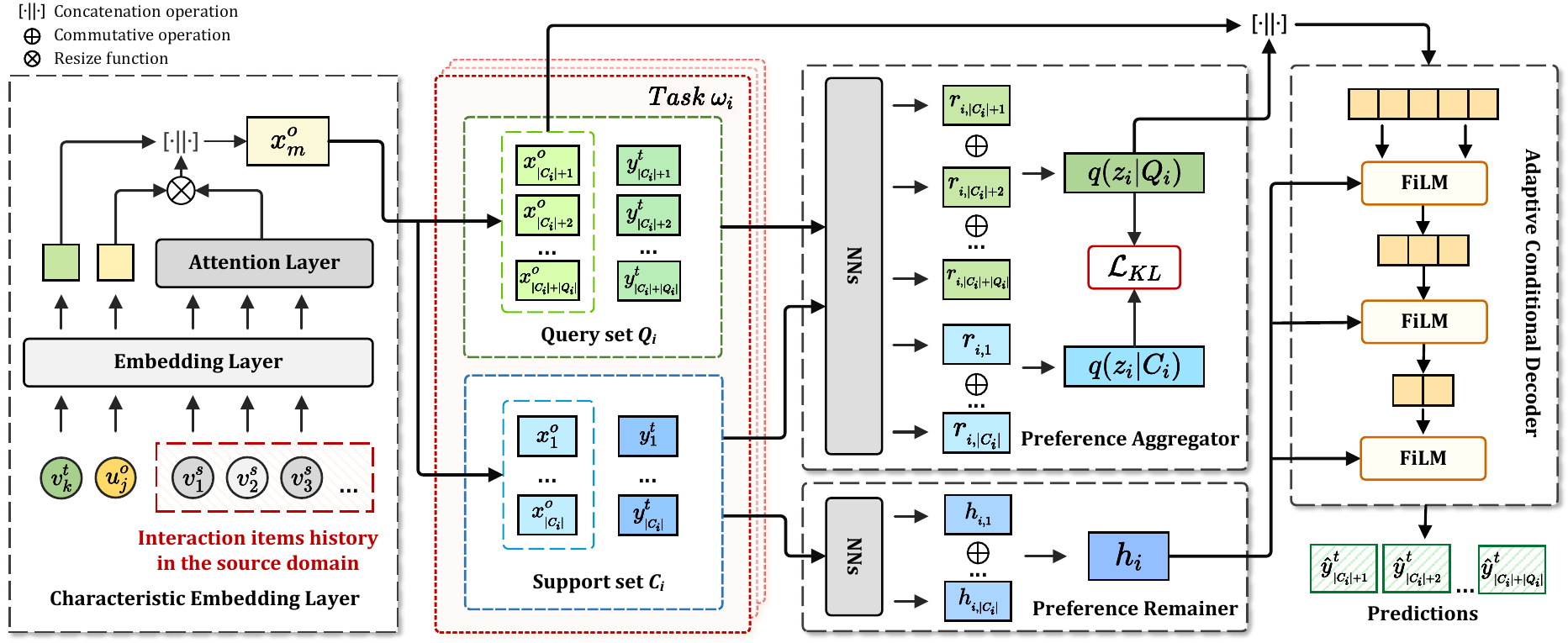}
		\caption{The framework of CDRNP in the training phase. The characteristic embedding layer is used to generate $\bm{x}_m^o$. Both of the $\mathcal{C}_i$ and $\mathcal{Q}_i$ are encoded to generate the variational prior and posterior. $\bm{h}_i$ learned from the preference remainer is used to modulate the decoder parameters. $\bm{z}_i$ sampled from $q(\bm{z}_i|\mathcal{Q}_i)$ is concatenated with $\bm{x}_m^o$ in $\mathcal{Q}_i$ to predict ${\hat{y}_m^t}$ for $\mathcal{Q}_i$ via decoder.}
		\label{fig:framework}
	\end{center}
\end{figure*}

\section{model}

This section introduces our model CDRNP implementing the NP principle. 
Fig. \ref{fig:framework} sketches the overall framework of CDRNP. 
Our model consists of three components: 
1) A characteristic embedding layer, which initializes user/item embeddings and generates  the rating-observed information representations for users.
2) A preference encoder, which contains a preference aggregator to achieve NP of the support and query set, and a preference remainer to enhance common preference representations from the support set.
3) An adaptive conditional decoder to incorporate features from the preference aggregator and preference remainer, and finally predict ratings of items in the target domain for the cold-start users.

\subsection{Characteristic Embedding Layer}\label{embedding}

In CDR, our goal is to predict $y^{t}_{m}$ in the $(x_{m}^{s\setminus o}, y^{t}_{m})$ pair for the cold-start users, given the $(x_{m}^{o}, y^{t}_{m})$ pairs of overlapping users.
Here $x_{m}$ (with superscript $s\setminus o$ or $o$) summarizes the rating-observed information $x_{m} \triangleq x_{jk} = (u_j, \mathcal{V}^{s}(u_j), v_k^t)$ for user $u_j$ to choose $v_k^t$ with the user's interactions $\mathcal{V}^{s}(u_j)$ in the source domain.
To achieve the above goal, we initialize the user/item representations in both domains, and then propose an attentive characteristic preference module to adaptively capture user's interactions $\mathcal{V}^{s}(u_j)$ in the source domain.
Finally, we fuse the user's information from the source domain with the candidate item $v_k^t$ in the target domain, so as to complete the rating-observed information $x_{m}$.

Specifically, we first initialize user embeddings as $\bm{u}_j \in \mathbb{R}^{d}$, and items embeddings as $\bm{v}_k^s \in \mathbb{R}^{d^2}$ in the source domain and $\bm{v}_k^t \in \mathbb{R}^{d^2}$ in the target domain\footnote{Here $d$ is the dimension of user embedding. As suggested in previous work~\cite{ptupcdr}, we let the dimension of item embedding be $d^2$ for the convenience of Eq.~(\ref{eq:embedding_fusion}).}.
Intuitively, various items that users interact in the source domain contribute differently to their preferences.
Therefore, we propose an attentive characteristic preference module over these item embeddings with attention mechanism~\cite{agm}:
\begin{equation}
\small
        \begin{split}
        &{a_k}=\bm{q}^{T}\text{ReLU}(\bm{W}_a\bm{v}_k^s), \\
        &{a_k^\prime}=\frac{exp(a_k)}{\sum_{v_n^s\in \mathcal{V}^{s}(u_j)}exp(a_n)},\\
        &\bm{c}_{u_j}=\sum_{v_k^s\in \mathcal{V}^{s}(u_j)}a_k^\prime \bm{v}_k^s,
        \end{split}
        \label{sec_math_eq:graph}
\end{equation}
where $\bm{q}\in\mathbb{R}^{d}$ and $\bm{W}_a\in\mathbb{R}^{d\times d}$ are learnable parameters. $a_k^\prime$ is the attention score of item $v_k^s$, differentiating the contribution of item $v_k^s$ to the user preference. 
Thereby, $\bm{c}_{u_j}\in \mathbb{R}^{d^2}$ summarizes the user-specific preference derived from the source domain.

To predict rating $y^{t}_{m}$ of candidate item $v_k^t$ for each user, we enhance the user feature (including $\bm{u}_j$ and $\bm{c}_{u_j}$) with the item embedding $\bm{v}_k^t$ in the target domain.
Specifically, we first fuse $\bm{u}_j$ and $\bm{c}_{u_j}$ to generate the user feature derived from the source domain, and then associate it with the candidate item:
\begin{equation}
\small
        \begin{split}
        {\hat{\bm{u}}_j} = \mathrm{Resize}(\bm{c}_{u_j}) \bm{u}_j, \quad
        {\bm{x}_m}=\left[\hat{\bm{u}}_j||\bm{v}_k^t\right],
        \end{split}
        \label{eq:embedding_fusion}
\end{equation}
where we resize $\bm{c}_{u_j}\in\mathbb{R}^{d^2}$ into $\mathbb{R}^{d\times d}$, which serves as a user-specific projection $\bm{W}_{u_j}$ in linear transformation, allowing better user feature fusion~\cite{ptupcdr}.
Besides, $\left[\cdot||\cdot\right]$ denotes concatenation operation.
Finally, we output $\bm{x}_m \in \mathbb{R}^{d}$ as the rating-observed information representation for the user $u_j$ with candidate item $v^t_k$.

Note that in the training phase, we leverage overlapping users in both $\mathcal{Q}_i$ and $\mathcal{C}_i$, so we can learn how to make prediction on $\mathcal{Q}_i$ by imitating $\mathcal{C}_i$.
In the testing phase, we make predictions for cold-start users in $\mathcal{Q}_i$, by imitating the overlapping users in $\mathcal{C}_i$.

\subsection{Preference Encoder}
This section introduces two modules: 1) a preference aggregator to capture preference correlations among users via NP; 2) a preference remainer to enhance original user preference features.

\subsubsection{Preference Aggregator}\label{sec:aggregator}
Recall that, we regard each task $\omega_i = \{(x_{m}, y^{t}_{m})|(x_{m}, y^{t}_{m})\in  \{\mathcal{C}_{i}\cup \mathcal{Q}_{i}\} \}$ decoupled from the same stochastic process $f_i$, and we approximate $f_i$ by representation $\bm{z}_i$ generated from neural networks.
Therefore, given $\mathcal{Q}_i$ and $\mathcal{C}_i$, we aim to generate the variational approximations $q(\bm{z}_i|\mathcal{Q}_i)$ and $q(\bm{z}_i|\mathcal{C}_i)$.

As suggested in previous NP studies~\cite{np}, we incorporate the $(x_{m}, y^{t}_{m})$ pair to approximate the stochastic process with neural networks, which can be formulated as:
\begin{equation}
\small
        \begin{split}
        \bm{r}_{i,m}=\mathrm{MLP}(\left[\bm{x}_m||\bm{y}_m^t\right];\theta),\quad (x_{m}, y^{t}_{m})\in \omega_i
        \end{split}
        \label{eq:np_fusion_xy}
\end{equation}
where $\bm{x}_m\in \mathbb{R}^d$, and $\bm{y}_m^{t}\in \mathbb{R}^1$ denotes a vector with one dimension of the scalar $y_m^t$.
Here we use $\mathrm{MLP}(\cdot;\theta)$ with three-layer neural networks parameterized by $\theta$ to fuse the features of $\bm{x}_m$ and $\bm{y}_m^t$.

Afterward, to obtain task representation $\bar{\bm{r}}_i$, we aggregate $\bm{r}_{i,m}$ with a commutative operation as in NP studies~\cite{np}. 
Taking $q(\bm{z}_i|\mathcal{C}_i)$ as an example, the aggregation process is as follows:
\begin{equation}
\small
        \begin{split}
        \bar{\bm{r}}_i &=\bm{r}_{i,1}\oplus \bm{r}_{i,2}\oplus \bm{r}_{i,3}\oplus\cdots \bm{r}_{i,|\mathcal{C}_i|-1}\oplus \bm{r}_{i,|\mathcal{C}_i|},\\
        &=\frac{1}{|\mathcal{C}_i|}\sum_{m=1}^{|\mathcal{C}_i|}\bm{r}_{i,m},
        \end{split}
        \label{eq:np_commutative}
\end{equation}
where $\oplus$ denotes a commutative operation, achieved by vector addition operation in our model for computational efficiency.
Note that $\bar{\bm{r}}_i$ is a permutation-invariance representation, which is a crucial property to ensure the $exchangeability$ condition of stochastic process~\cite{stochastic}. 
Such a design reduces the time complexity to $O(n+m)$, where $n$ and $m$ are the number of users in $\mathcal{C}_i$ and $\mathcal{Q}_i$, respectively.

To derive the random variable $\bm{z}_i$ of the stochastic process, as widely studied in variantional frameworks~\cite{aevb}, we employ Gaussian distribution to represent $\bm{z}_i$, i.e., imposing $\bm{z}_i\sim \mathcal{N}(\bm{\mu}_i,\bm{\sigma}_i^2)$.
Therefore, we generate $\bm{z}_i$ from the task representation $\bar{\bm{r}}_i$ as:
\begin{equation}
\small
        \begin{split}
        \hat{\bm{r}}_i=\text{ReLU}&(\bm{W}_r \bar{\bm{r}}_i),\\
        \bm{\mu}_i=\bm{W}_\mu \hat{\bm{r}}_i,
        \quad &\log\bm{\sigma}_i=\bm{W}_\sigma \hat{\bm{r}}_i,
        \end{split}
        \label{sec_math_eq:graph}
\end{equation}
where $\bm{W}_r\in\mathbb{R}^{d\times d}$, $\bm{W}_\mu\in\mathbb{R}^{d\times d}$, $\bm{W}_\sigma\in\mathbb{R}^{d\times d}$ are learnable parameters.
However, the random variable is intractable in training neural networks, we employ the well-known reparameterization trick~\cite{aevb} to sample data from probability distribution, thus the gradients can be smoothly backpropagated:
\begin{equation}
\small
        \begin{split}
        \bm{z}_i&=\bm{\mu}_i + \bm{\epsilon}\odot\bm{\sigma}_i, \quad \bm{\epsilon} \sim \mathcal{N}(0,\bm{I}),
        \end{split}
        \label{eq:z}
\end{equation}
where $\odot$ denotes the element-wise product operation, $\bm{\epsilon}$ is the Gaussian noise.
In this way, we can sample $\bm{z}_i$ from both $q(\bm{z}_i|\mathcal{Q}_i)$ and $q(\bm{z}_i|\mathcal{C}_i)$ in $\mathcal{Q}_i$ and $\mathcal{C}_i$, respectively.
Besides, by formulating all tasks involving users' interactions into the same stochastic process, we capture preference correlations among users for the CDR problem.

\subsubsection{Preference Remainer}\label{sec:preference_remainer}
To further enhance the preference correlations among users in different layer-depths, we devise a preference remainer to retain the original overlapping users preference features from the support set $\mathcal{C}_i$.
Specifically, we first generates user preference representation $\bm{h}_{i,m}$ by fusing rating-observed information $x_m^o$ and the item rating $y_m^t$ in the target domain, observed in the given support set $\mathcal{C}_i$ as:
\begin{equation}
\small
        \begin{split}
        \bm{h}_{i,m}=\mathrm{MLP}(\left[\bm{x}_m^o||\bm{y}_m^t\right];\phi), \quad (x_{m}^{o}, y^{t}_{m})\in \mathcal{C}_i
        \end{split}
        \label{sec_math_eq:graph}
\end{equation}
where $\mathrm{MLP}(\cdot;\phi)$ denotes three-layer neural networks parameterized by $\phi$ and $\left[\cdot||\cdot\right]$ denotes the concatenation operation.
Then, we derive the common preference representation ${\bm{h}}_{i}$ by aggregating each user preference representation $\bm{h}_{i,m}$ as follows: 
\begin{equation}
\small
        \begin{split}
        {\bm{h}}_i&=\bm{h}_{i,1}\oplus \bm{h}_{i,2}\oplus \bm{h}_{i,3}\oplus\cdots 
        \bm{h}_{i,|\mathcal{C}_i|-1}\oplus \bm{h}_{i,|\mathcal{C}_i|},\\
        &=\frac{1}{|\mathcal{C}_i|}\sum_{m=1}^{|\mathcal{C}_i|}\bm{h}_{i,m},
        \end{split}
        \label{sec_math_eq:graph}
\end{equation}
Here we use the same operation in Eq.~(\ref{eq:np_commutative}) for convenience.
In this way, we derive the common preference representations from the support set $\mathcal{C}_i$, enriching the preference correlations among users for making prediction in the query set $\mathcal{Q}_i$.

\subsection{Adaptive Conditional Decoder}\label{decoder}
To make prediction for rating $y^{t}_{m}$ given $x_{m}$ in $\mathcal{Q}_i$, we have obtained the preference information from $\mathcal{C}_i$, including $\bm{z}_i$ from the preference aggregator and ${\bm{h}}_i$ from the preference remainer.
To better incorporate $\bm{z}_i$ and ${\bm{h}}_i$ for predicting item ratings, we propose an adaptive conditional decoder based on the well-designed FiLM~\cite{film,ada}, which can fully modulate the two features in different layer-depths for decoding.

Specifically, the adaptive conditional decoder learns the conditional likelihood $p(y_m^t|x_m,\bm{z}_i)$ (in Eq.(\ref{eq:np_lowerbound_meta})) with the $x_m$ and $\bm{z}_i$ (in Eq.~(\ref{eq:z})) as inputs.
We first generate two modulation coefficients $\bm{\gamma}_i$ and $\bm{\beta}_i$ over the representation $\bm{h}_i$, and then the adaptive conditional decoder can be formulated as:
\begin{equation}
\small
        \begin{split}
        &\bm{\gamma}_i^l=\text{tanh}(\bm{h}_i;\bm{W}_{\gamma}^l),\quad \bm{\beta}_i^l=\text{tanh}(\bm{h}_i;\bm{W}_{\beta}^l),\\
        \bm{g}^{0} = &\left[\bm{x}_m||\bm{z}_i\right],\quad \bm{g}^{l+1} = \text{ReLU}(\bm{\gamma}_i^l\odot(\bm{W}_g^l \bm{g}^{l}+\bm{b}_g^l)+\bm{\beta}_i^l), \\
        \end{split}
        \label{sec_math_eq:graph}
\end{equation}
where $\bm{g}^{l}$ is the input of $l$-th layer and $\left[\cdot||\cdot\right]$ denotes the concatenation operation. 
We use tanh to enforce the modulation coefficients in range [-1,1]. 
In this way, $\bm{\gamma}_i^l$ controls the gain from $\bm{g}^{l}$, while $\bm{\beta}_i$ remains the features of $\bm{h}_i$.
By stacking multiple layers, the adaptive conditional decoder fully captures preference correlations among overlapping and cold-start users in different layer-depths, and finally output the rating $\hat{y}_m^t$ of user $u_j$ on item $v_k^t$ at the last layer.

\subsection{Loss Function}
To train our model, the loss function consists of two terms: the desired log-likelihood to achieve CDR, and the KL divergence to impose regularization. 
Particularly, we rephrase the likelihood $\mathcal{L}_{rec}$ in Eq.~(\ref{eq:np_loss}) as a regression-based loss function by MSE loss:
\begin{equation}
\small
        \begin{split}
        \mathcal{L}_{rec}&=-E_{q(z_i|\mathcal{Q}_i)}\log p(y_{1:|\mathcal{Q}_i|}^t|x_{1:|\mathcal{Q}_i|}^o,z_i),\\
        &\propto\frac{1}{|\mathcal{Q}_i|}\sum_{m=1}^{|\mathcal{Q}_i|}(y_m^t-\hat{y}_m^t)^2,
        \end{split}
        \label{sec_math_eq:graph}
\end{equation}
where we use overlapping users $x_{1:|\mathcal{Q}_i|}^o$ to replace $x_{1:|\mathcal{Q}_i|}$.
Besides, we impose $\mathcal{L}_{KL}$ for the $consistency$ condition in the approximated stochastic process, which practically minimizes the KL divergence in Eq.~(\ref{eq:np_loss}) between the approximate posterior of the query set $\mathcal{Q}_i$ and conditional prior of the support set $\mathcal{C}_i$:
\begin{equation}
\small
        \begin{split}
        \mathcal{L}_{KL}=\mathrm{KL}(q(z_i|\mathcal{Q}_i)||q(z_i|\mathcal{C}_i)),
        \end{split}
        \label{sec_math_eq:graph}
\end{equation}
In summary, the overall loss function is defined as follows:
\begin{equation}\label{loss}
\small
        \begin{split}
        \mathcal{L}=\mathcal{L}_{rec}+\lambda \mathcal{L}_{KL},
        \end{split}
\end{equation}
where $\lambda \in [0, 1]$ is the hyper-parameter to balance losses.

Note that we use overlapping users in the training phase, CDRNP first learns on support set $\mathcal{C}_i$ and then updates according to the prediction loss over query set $\mathcal{Q}_i$. 
In the testing phase, we randomly removed all the ground-truth rating $y_m^t$ in $\mathcal{Q}_i$ and regard them as the test users (cold-start users). 
CDRNP can predict the preference of the cold-start users in $\mathcal{Q}_i$ via a new task $\omega_i$ with only overlapping users in $\mathcal{C}_i$ available and the global information learned from the training phase.
Thus given $\mathcal{Q}_i$ and $\mathcal{C}_i$, the decoder takes $z_i$ sampled from $q(z_i|\mathcal{C}_i)$ and $x_m^{s\setminus o}$ in $\mathcal{Q}_i$ as inputs to predict ${\hat{y}_m^t}$ for $\mathcal{Q}_i$.

\section{Experiments}
In this section, we compare our proposed model with previous models, and conduct experiments to evaluate our model.

\subsection{Datasets}

Following existing methods~\cite{sscdr,catn,tmcdr} on CDR, we evaluate CDRNP on a real-world recommendation dataset, namely the Amazon review dataset.
Specifically, we use the Amazon-5scores dataset where each user or item has at least five ratings.

Following \cite{sscdr,catn}, we select three popular categories out of 24 categories for CDRNP: movies$\_$and$\_$tv (Movie), cds$\_$and$\_$vinyl (Music), and books (Book). We define three CDR scenarios: Movie-domain$\to$Music-domain, Book-domain$\to$Movie-domain, and Book-domain$\to$Music-domain. While many existing works only evaluate a subset of data, we conduct experiments using all available data, which more realistically simulates the real-world. Table \ref{sec_exp_tab:dataset} summarizes the detailed information of three CDR scenarios.

\begin{table}[t]
\centering
\footnotesize
\caption{Statistics of Three CDR scenarios (\textit{Overlap} denotes the number of overlapping users).}
\setlength{\tabcolsep}{1.3pt}
{
\begin{tabular*}{0.47 \textwidth}
{@{\extracolsep{\fill}}@{}ccc||ccc||cc@{}}
\toprule
\multirow{2}{*}{\bf Scenarios}  &\multicolumn{2}{c}{\bf \#Items} & \multicolumn{3}{c}{\bf \#Users} & \multicolumn{2}{c}{\bf \#Ratings}\\
\cmidrule(r){2-3}\cmidrule(r){4-6}\cmidrule(r){7-8}&
Source & Target & Overlap &  Source & Target & Source &  Target  \\
\midrule
\bf Movie $\to$ \bf Music  &50,052  &64,443 &18,031   &123,960   &75,258  &1,697,533 
 &1,097,592     \\ 

\bf Book $\to$ \bf Movie  &367,982  &50,052 &37,388   &603,668   &123,960  &8,898,041  &1,697,533    \\ 

\bf Book $\to$ \bf Music  &367,982  &64,443 &16,738   &603,668   &75,258  &8,898,041  &1,097,592    \\ 
 
\bottomrule
\end{tabular*}
}
\label{sec_exp_tab:dataset}
\end{table}

\begin{table*}[t]
\footnotesize
\centering
\caption{Performance comparison of CDR to cold-start users in three CDR scenarios.}
\label{mainexperiment}
\setlength\tabcolsep{2.0pt}
\begin{tabular*}{1 \textwidth}
{@{\extracolsep{\fill}}@{}lcccccc|cccccc|cccccc@{}}
\toprule
\multirow{3}{*}{\renewcommand{\arraystretch}{0.8} \bf Methods}  &
\multicolumn{6}{c}{\bf Movie-domain$\to$Music-domain} & \multicolumn{6}{c}{\bf Book-domain$\to$Movie-domain} & \multicolumn{6}{c}{\bf Book-domain$\to$Music-domain}    \\
\cmidrule(r){2-7}\cmidrule(r){8-13}\cmidrule(r){14-19}&
\multicolumn{2}{c}{ 20$\%$} & \multicolumn{2}{c}{ 50$\%$} & \multicolumn{2}{c}{ 80$\%$} &  \multicolumn{2}{c}{ 20$\%$} & \multicolumn{2}{c}{ 50$\%$} & \multicolumn{2}{c}{ 80$\%$} &  \multicolumn{2}{c}{ 20$\%$} & \multicolumn{2}{c}{ 50$\%$} & \multicolumn{2}{c}{ 80$\%$} \\
\cmidrule(r){2-3}\cmidrule(r){4-5}\cmidrule(r){6-7}
\cmidrule(r){8-9}\cmidrule(r){10-11}\cmidrule(r){12-13}
\cmidrule(r){14-15}\cmidrule(r){16-17}\cmidrule(r){18-19}&
MAE & RMSE & MAE & RMSE & MAE & RMSE & MAE & RMSE & MAE & RMSE & MAE & RMSE & MAE & RMSE & MAE & RMSE & MAE & RMSE \\
\midrule
TGT  &   4.4803   &   5.1580   &   4.4989   &   5.1736   &   4.5020   &   5.1891   & 
 4.1831   &   4.7536   &   4.2288   &   4.7920   &   4.2123   &   4.8149   &   4.4873   &   5.1672   &   4.5073   &   5.1727   &   4.5204   &   5.2308  \\

CMF    &  1.5209 &   2.0158    &   1.6893    &    2.2271 &   2.4186  &   3.0936 & 
 1.3632   &   1.7918 &  1.5813   &    2.0886  &    2.1577    &    2.6777  &  1.8284  &  2.3829  &  2.1282  &  2.7275  &  3.0130  &  3.6948  \\
\midrule
DCDCSR &  1.4918 &   1.9210    &   1.8144    &    2.3439 &   2.7194  &   3.3065 & 
 1.3971   &   1.7346 &  1.6731   &    2.0551  &    2.3618    &    2.7702  &  1.8411  &  2.2955  &  2.1736  &  2.6771  &  3.1405  &  3.5842  \\
SSCDR   &    1.3017 &   1.6579    &   1.3762    &    1.7477 &   1.5046  &   1.9229 & 
 1.2390   &   1.6526 &  1.2137   &    1.5602  &    1.3172    &    1.7024  &  1.5414  &  1.9283  &  1.4739  &  1.8441  &   1.6414  &  2.1403  \\
EMCDR   &  1.2350 &   1.5515 &  1.3277 &  1.6644 &  1.5008  &  1.8771  &
 1.1162 &  1.4120   &  1.1832 &  1.4981 &  1.3156 &  1.6433  &  1.3524  &  1.6737  &  1.4723  &  1.8000  &  1.7191  &  2.1119  \\

PTUPCDR   &   \underline{1.1504} &   \underline{1.5195} &  \underline{1.2804} &  \underline{1.6380} &  \underline{1.4049}  &  \underline{1.8234}  &
 \underline{0.9970} &  \underline{1.3317}   &  \underline{1.0894} &  \underline{1.4395} &  \underline{1.1999} &  \underline{1.5916}  &  \underline{1.2286}  &  \underline{1.6085}  &  \underline{1.3764}  &  \underline{1.7447}  &  \underline{1.5784}  &  \underline{2.0510}  \\
\midrule
CDRNP   &  \textbf{0.7974} &  \textbf{1.0638}   &  \textbf{0.7969} &  \textbf{1.0589} &  \textbf{0.8280} &  \textbf{1.0758} &
  \textbf{0.8846} &  \textbf{1.1327}   &  \textbf{0.8946} &  \textbf{1.1450} &  \textbf{0.8970} &  \textbf{1.1576}  &  \textbf{0.7453}  &  \textbf{0.9914}  &  \textbf{0.7629}  &  \textbf{1.0111}  &  \textbf{0.7787}  &  \textbf{1.0373}  \\
Improve & {30.68\%}  &  {29.99\%}  &  {37.76\%}  &  {35.35\%}  &  {41.06\%}  &  {41.00\%}  &  {11.27\%}  &  {14.94\%}  &  {17.88\%}  &  {20.46\%}  &  {25.24\%}  &  {27.27\%}  &  {39.34\%}  &  {38.36\%}  &  {44.57\%}  &  {42.05\%}  &  {50.67\%}  &  {49.42\%} \\
\bottomrule
\end{tabular*}
\end{table*}

\subsection{Experimental Setting}
\subsubsection{Evaluation Metrics} 
\label{sec:leave_one_out}
Following previous work \cite{emcdr,catn}, we use mean absolute error (MAE) and root mean square error (RMSE) as evaluation metrics to assess the recommendation performance. For each metric, we report the average results over five random runs for all users. Following previous meta-learning recommenders \cite{mlcdr,mamo,melu}, we only predict the score of each user on candidate items in the query set $\mathcal{Q}_i$.

\subsubsection{Compared Baselines} 
For a fair comparison, we choose the following methods as baselines:
(1) \textbf{TGT} represents the target MF model, which only uses data from the target domain for training and ignores user interactions in the source domain.
(2) \textbf{CMF} \cite{cmf} leverages the source domain interactions as additional features for overlapping users to achieve performance improvement in the target domain. 
(3) \textbf{EMCDR} \cite{emcdr} follows the embedding and mapping idea for CDR, learning specific user representations in each domain through a mapping function, which is then utilized to transfer user representations from the source domain to the target domain.
(4) \textbf{DCDCSR} \cite{dfcdr} is a CDR model based on matrix factorization and DNNs. It generates latent vectors for users and items using MF model, then uses DNNs to map latent vectors across domains.
(5) \textbf{SSCDR} \cite{sscdr} proposes a CDR framework based on semi-supervised mapping. It trains the mapping function by exploiting overlapping users and all items. Moreover, it predicts the latent vectors of cold-start users by aggregating neighborhood information.
(6) \textbf{PTUPCDR} \cite{ptupcdr} is a meta-learning-based CDR method that generates personalized mapping functions for each user by learning a meta-network, transferring the user's preference from the source domain to the target domain.

\subsubsection{Implementation and Hyperparameter Setting}
The source codes for all baselines have been released, and we modify the data input and evaluation sections to fit our experimental settings. To ensure a fair comparison, we set the same hyper-parameter settings for all methods. For each method, we tune the dimension of embedding in $\left\{{32, 64, 128, 256}\right\}$. All models are trained for 10 epochs to achieve convergence. We use the same fully connected layer to facilitate comparison for CDRNP and other baselines. The stacking layers for preference aggregator, preference remainer and adaptive conditional decoder in our model are three, for attentive characteristic preference module is two. The learning rate is chosen from $\left\{{0.001, 0.005, 0.01, 0.02, 0.1}\right\}$. 
The hidden layer size for each layer is chosen from $\left\{{8, 16, 32, 64, 128}\right\}$, and the hyper-parameter $\lambda$ in Eq.~(\ref{loss}) is chosen from $\left\{{0.1, 0.3, 0.5, 0.7, 0.9}\right\}$.
In our experiments, we set the proportion of the testing dataset users $\alpha$ to be 20$\%$, 50$\%$, and 80$\%$ of the total overlapping users, respectively. We consider the sequential timestamps to avoid information leakage.

\subsection{Performance Comparisons on CDR}\label{performance comparisons}
This section presents the experimental results and detailed analyses of CDRNP in three CDR scenarios and in-depth discussion of the CDRNP bi-directional recommendation performance for cold-start users in both the source and target domains.
\subsubsection{Performance Comparisons}\label{sec:main_comparison}
Table \ref{mainexperiment} shows the experimental performance of all methods in three CDR scenarios. 
Note that the size of $\mathcal{C}_i$ is set to 40, and the length of interaction items history $\mathcal{V}^{s}(u_j)$ is set to 20. 
Based on the experimental results, we can draw the following conclusions:
(1) {\textit{CDRNP consistently exhibits the best performance in all CDR scenarios.}} Specifically, under the experimental setting of $\alpha$ value at 20$\%$, CDRNP outperforms the existing SOTA (i.e., PTUPCDR) in MAE by 30.68$\%$, 11.27$\%$ and 39.34$\%$ in three CDR scenarios.
(2) {\textit{CDRNP demonstrates outstanding performance and robustness even when the training dataset is limited.}} For example, for Movie domain$\to$Music domain, with $\alpha$ value set to 20\%, 50\%, and 80\%, CDRNP achieves improvements of 30.68\%, 37.76\%, and 41.06\%, respectively, compared to the state-of-the-art meta-learning method. This indicates that our NP framework can effectively adapt to CDR even with limited trainable overlapping users.
(3) {\textit{CDRNP achieves competitive performance compared with all EMCDR-based and meta-learning-based baselines.}}
The results highlight the necessity of capturing the preference correlations among overlapping and cold-start users.

\begin{table}[t]
\centering
\footnotesize
\caption{Performance of bi-directional recommendation.}
\label{bidirectional}
\setlength\tabcolsep{2pt}
{
\begin{tabular*}{0.47 \textwidth}
{@{\extracolsep{\fill}}@{}ccc|cc@{}}
\toprule
\multirow{2}{*}{\bf \text{Model}}  &  
\multicolumn{2}{c}{$Movie\to Music$} & \multicolumn{2}{c}{$Music\to Movie$} \\ 
\cmidrule(r){2-3}\cmidrule(r){4-5}&
MAE & RMSE & MAE & RMSE \\
\midrule
$\text{EMCDR}_{Movie\to Music}$ &   1.6573 & 2.0170  &  4.1198 & 4.5604 \\
$\text{EMCDR}_{Music\to Movie}$ &   4.3590 &  4.9620 &  1.1773 & 1.5165 \\
\midrule
$\text{PTUPCDR}_{Movie\to Music}$   &  1.2859   &  1.7384 &   4.4992 & 5.3533  \\
$\text{PTUPCDR}_{Music\to Movie}$ &  4.3781   &  4.9627 &   1.0954 & 1.4161\\
\midrule
CDRNP  & 0.9140   &  1.1257 &  0.8817 & 1.1383 \\

\bottomrule
\end{tabular*}
}
\end{table}

\subsubsection{Bi-directional Recommendation Performance}\label{sec:bi-recommendation}
We evaluate the bi-directional recommendation performance of CDRNP to cold-start users in both the source and target domain with EMCDR and PTUPCDR. Specifically, we first split 50$\%$ of the overlapping users for training EMCDR and PTUPCDR in the Movie-domain$\to$Music-domain. To ensure fairness, we split 25$\%$ of overlapping users for testing EMCDR and PTUPCDR in Movie-domain$\to$Music-domain and the remained 25$\%$ of overlapping users for testing in Music-domain$\to$Movie-domain, respectively. We take the same operation to train CDRNP, EMCDR and PTUPCDR in the Music-domain$\to$Movie-domain, where testing them in the Music-domain$\to$Movie-domain and Movie-domain$\to$Music-domain. The empirical results are presented in Figure \ref{bidirectional}, from which we can draw the following conclusions:
(1) {\textit{CDRNP achieved the best bi-directional recommendation performance to cold-start users in both the source and target domains.}} CDRNP only needs to be trained in the Music-domain$\to$Movie-domain to achieve outstanding bi-directional recommendation performance for cold-start users in both the Music-domain and Movie-domain.
(2) {\textit{In comparison with EMCDR and PTUPCDR, CDRNP can achieve outstanding CDR performance to cold-start users in the target domain.}} This highlights the necessity of utilizing explicit user-item interactions from both domains.

\begin{table}[t]
\renewcommand\arraystretch{1}
\centering
\footnotesize
\caption{Ablation Study on three CDR scenarios.}
\label{ablation}
\setlength\tabcolsep{2pt}{
\begin{tabular*}{0.47 \textwidth}
{@{\extracolsep{\fill}}@{}cc|ccc@{}}
\toprule
\bf \text {Model} & \bf \text {Metric} & \bf \text {Scenario 1}  & \bf \text {Scenario 2} & \bf \text {Scenario 3}  \\
\midrule
\multirow{2}{*}{CDRNP(\textit{w}/\textit{o} PRM+ACP)} & \text {MAE} &   0.8597 &  0.9446 &  0.7940 \\
&\text {RMSE} &   1.0908 &  1.1645 &  1.0214 \\
\midrule
\multirow{2}{*}{CDRNP(\textit{w}/\textit{o} PRM)}   &   \text {MAE}  &  0.8259   &  0.9002 &   0.7628  \\
&\text {RMSE} &  1.0735   &  1.1409 &   1.0019\\
\midrule
\multirow{2}{*}{CDRNP} & \text {MAE} &   0.7974   &  0.8846 &  0.7453    \\
&\text {RMSE} & 1.0638   &  1.1327 &  0.9914 \\

\bottomrule
\end{tabular*}
}
\end{table}

\subsection{Ablation Study}\label{sec:ablation}
In this section, we conduct an ablation study to test the effectiveness of each module in CDRNP. 
We first remove the Preference Remainer (PRM) module (FiLM layer number 0) from CDRNP, and then remove the Attentive Characteristic Preference (ACP) module based on that. 
We evaluate the intact CDRNP (FiLM layer number 3). 
We utilize the $\alpha$ value of 20{\%} for all CDR scenarios. 
From the results shown in Table \ref{ablation}, we have several findings:
(1) {\textit{The CDR performance of CDRNP(w/o PRM) significantly outperforms that of CDRNP(w/o PRM+ACP).}} The results indicate the effectiveness of transferring the preference of users from the source domain to the target domain, 
and highlight the necessity of using the ACP module to utilize auxiliary information from the source domain for CDR.
(2) {\textit{The improvement of CDRNP over CDRNP (w/o PRM) is significant.}} The results demonstrate the importance of capturing the preference correlations among users in different layer-depths.

\begin{figure}[t]
\setlength{\abovecaptionskip}{0.cm}
	\begin{center}
        \subfigure
        {\begin{minipage}[b]{.32\linewidth}
        \centering
        \includegraphics[scale=0.23]{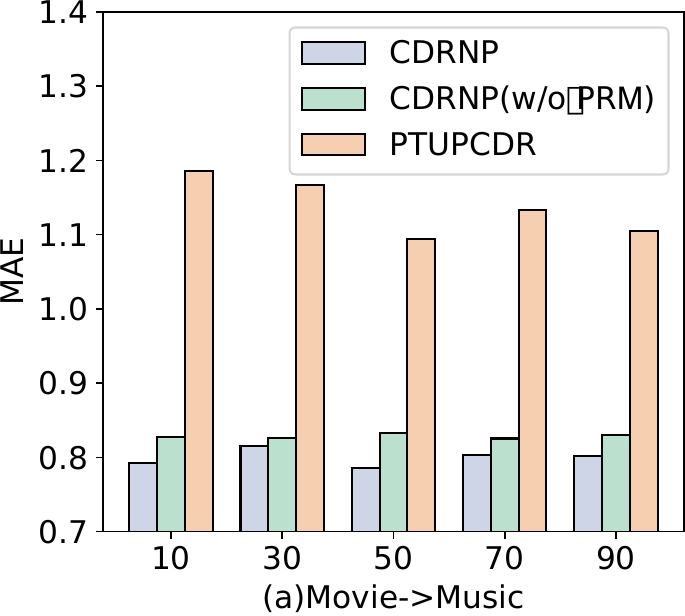}
        \end{minipage}}
        \subfigure
        {\begin{minipage}[b]{.32\linewidth}
        \centering
        \includegraphics[scale=0.23]{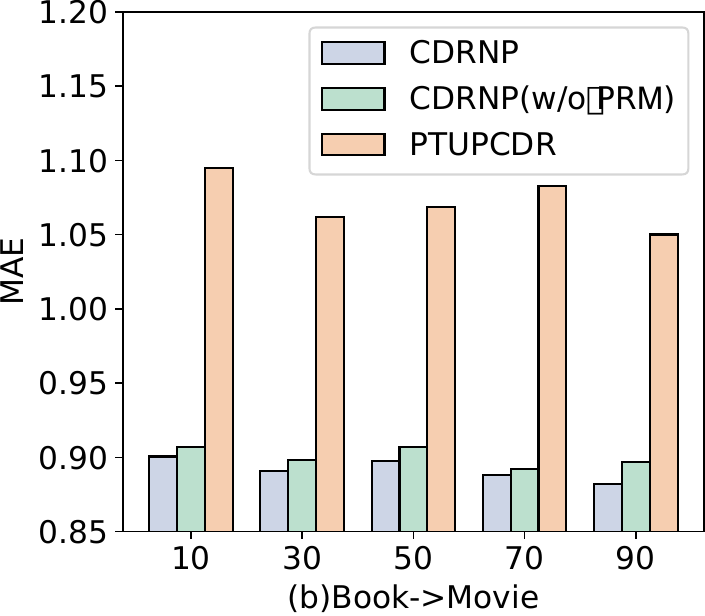}
        \end{minipage}}
        \subfigure
        {\begin{minipage}[b]{.32\linewidth}
        \centering
        \includegraphics[scale=0.23]{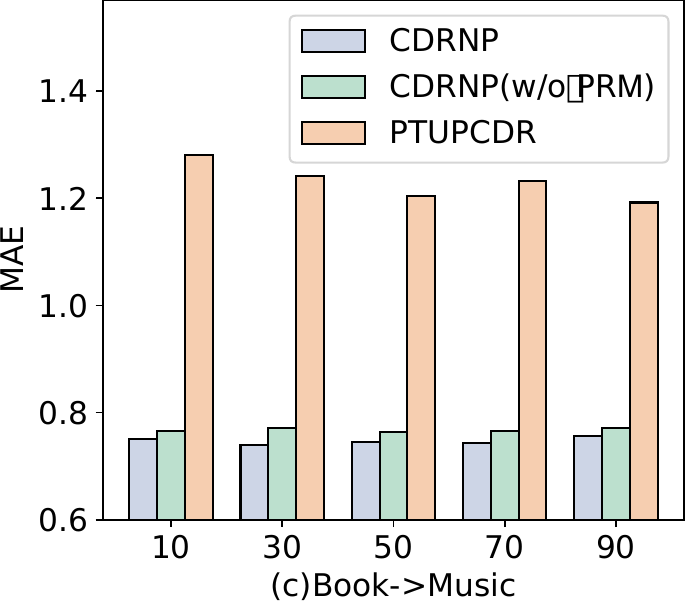}
        \end{minipage}}
        
	\caption{Performance comparison of CDR to cold-start users with different lengths of interaction items history $\mathcal{V}^{s}(u_j)$.}
	\label{history}
	\end{center}
\end{figure}

\begin{figure}[t]
\setlength{\abovecaptionskip}{0.cm}
	\begin{center}
        \subfigure
        {\begin{minipage}[b]{.32\linewidth}
        \centering
        \includegraphics[scale=0.225]{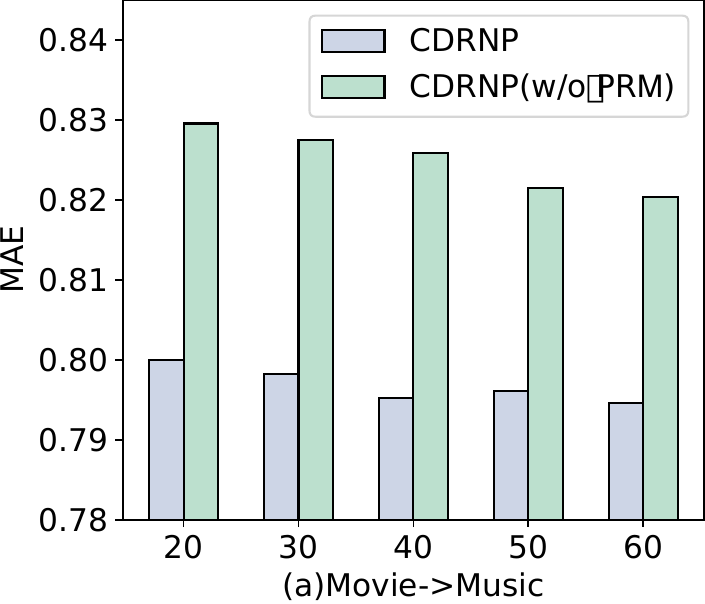}
        \end{minipage}}
        \subfigure
        {\begin{minipage}[b]{.32\linewidth}
        \centering
        \includegraphics[scale=0.225]{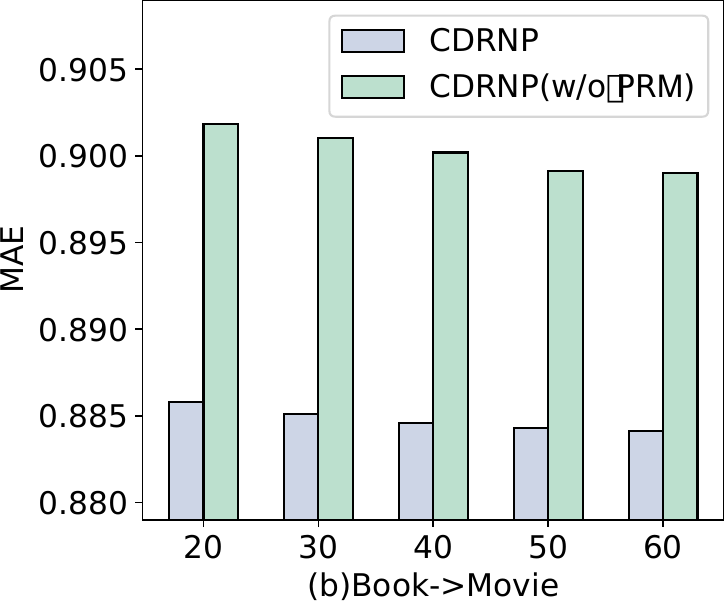}
        \end{minipage}}
        \subfigure
        {\begin{minipage}[b]{.32\linewidth}
        \centering
        \includegraphics[scale=0.225]{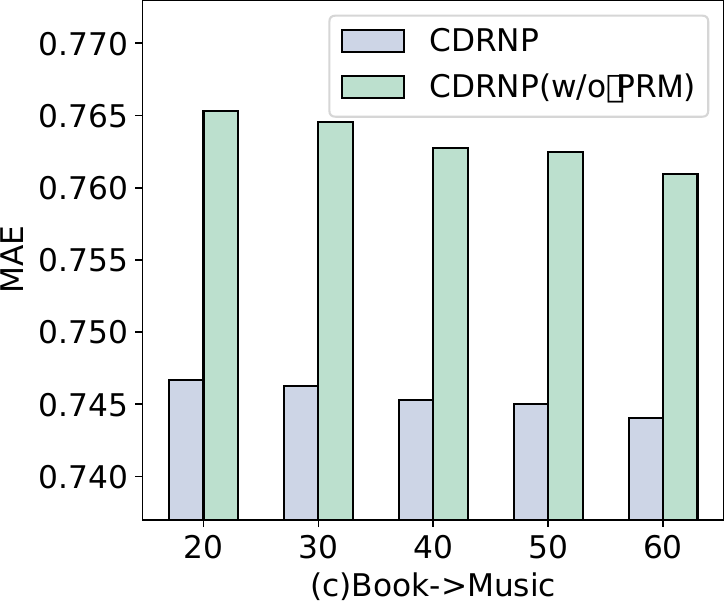}
        \end{minipage}}
        
	\caption{Performance comparison of CDR for cold-start users with different lengths of support set $\mathcal{C}_i$.}
	\label{support}
	\end{center}
\vspace{-1em}
\end{figure}

\subsection{Impact of $|\mathcal{V}^{s}(u_j)|$ and $|\mathcal{C}_i|$}\label{RQ2}

In this section, we modify the length of interaction items history $\mathcal{V}^{s}(u_j)$ and the number of users in $\mathcal{C}_i$ separately in three CDR scenarios with $\alpha$ value of 20\% to test the effectiveness of CDRNP.

\subsubsection{Impact of $|\mathcal{V}^{s}(u_j)|$}\label{sec:sum} 
In this section, $|\mathcal{V}^{s}(u_j)|$ is chosen from $\left\{{10, 30, 50, 70, 90}\right\}$.  
From the results shown in Figure \ref{history}, we can draw the following conclusions:
(1) {\textit{Despite a short interaction items history, CDRNP exhibits exceptional recommendation performance.}} Our model shows robust performance with regard to the length of interaction items history in the source domain.
(2) {\textit{CDRNP consistently outperforms CDRNP(\textit{w}/\textit{o} PRM) and PTUPCDR when the interaction items history is reduced.}} This indicates that our model is more effective in generating accurate recommendations using limited interaction items history in the source domain.
(3) {\textit{Our model is less affected by the length of the interaction items history.}} It is possibly due to its ability to model preference correlations among the overlapping and cold-start users and user-specific preference from both domains, resulting in complete preferences of users. In contrast, PTUPCDR is more sensitive to the change of $|\mathcal{V}^{s}(u_j)|$.

\subsubsection{Impact of $|\mathcal{C}_i|$}\label{sec:nsi}
In this section, we set $|\mathcal{C}_i|$ to $\left\{{20, 30, 40, 50, 60}\right\}$. 
From the results shown in Figure \ref{support}, we can draw the following conclusions:
(1) {\textit{Both CDRNP and CDRNP (\textit{w}/\textit{o} PRM) maintain robust performance when the number of users in the support set $\mathcal{C}_i$ is reduced.}} This indicates that our model is less affected by the reduction of users in $\mathcal{C}_i$. The stability may be attributed to the fact that CDRNP applies a random function to the constructions of $\omega_i$ to simulate different function realisations, making the performance of CDRNP more stable.
(2) {\textit{CDRNP outperforms CDRNP (\textit{w}/\textit{o} PRM) in all experimental settings.}} The results further emphasize the importance of capturing preference correlations among the overlapping and cold-start users in different layer-depths.

\begin{figure}[t]
\setlength{\abovecaptionskip}{0.cm}
	\begin{center}
        \subfigure
        {\begin{minipage}[b]{.45\linewidth}
        \centering
        \includegraphics[scale=0.27]{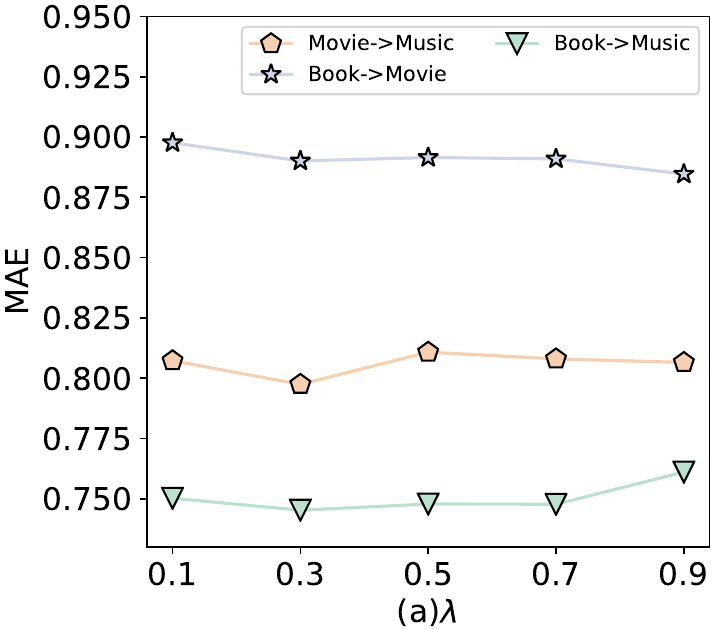}
        \end{minipage}}
        \subfigure
        {\begin{minipage}[b]{.45\linewidth}
        \centering
        \includegraphics[scale=0.27]{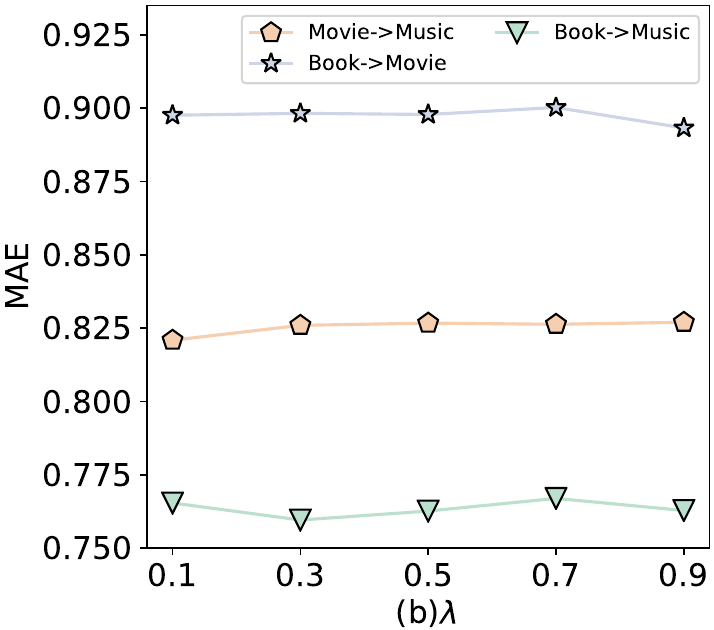}
        \end{minipage}}

	\caption{Effect of the training loss factor $\lambda$.}
	\label{lamda}
	\end{center}
\vspace{-0.9em}
\end{figure}

\subsection{Hyper-parameter Analysis}\label{sec:hyper-parameter}
In this section, we investigate the sensitivity of CDRNP to the hyperparameter $\lambda$ in Eq. (\ref{loss}). This experiment is conducted in three CDR scenarios with $\alpha$ value of 20{\%}. Figure \ref{lamda}(a) shows results of CDRNP, and Figure \ref{lamda}(b) for CDRNP(\textit{w}/\textit{o} PRM). From these results, we can draw the following conclusions:
(1) {\textit{We find $\lambda$ can be set as a larger value for larger interaction scenario, such as the Book-domain$\to$Movie-domain.}} The reason might be that the larger interaction scenario contains more complex preference correlations among the overlapping and cold-start users, which should be captured using a larger $\lambda$. In contrast, for the other two scenarios, a smaller $\lambda$ can achieve good recommendation performance. 
(2) {\textit{The experimental results of our model are stable, which verifies the robustness of our model to changes in $\lambda$.}} Even in the worst case of $\lambda$, our model still outperforms other baselines in Table \ref{mainexperiment}.

\section{Related Works}
\textbf{Cross-domain recommendation} leverages observed interactions from source domian to improve the recommendation performance in the target domian.
Following the collaborative filtering (CF) paradigm, a significant amount of CDR methods~\cite{BPR,cmf,ncf,ngcf} have been proposed to make effective recommender systems. Although these CF-based methods achieve outstanding performance, they still suffer from the cold-start and data sparsity issues. 
As our CDRNP aims to solve the previous issue, we only focus on CDR methods which recommend items to cold-start users.

To mitigate the cold-start issue, several CDR methods have been proposed to establish correlations between the source and target domains.
Specifically, EMCDR \cite{emcdr} follows the embedding and mapping paradigm, which learns user representations in each domain with a mapping function. DCDCSR \cite{dfcdr} utilizes MF and DNN to generate benchmark factors from
source and target domains. SSCDR \cite{sscdr} prposes a semi-supervised mapping function by exploiting overlapping users and all items. PTUPCDR \cite{ptupcdr} and TMCDR \cite{tmcdr} follow the MAML \cite{mamlfa} framework to learn a meta network to substitute the mapping function. 
CATN \cite{catn} proposes a model based on user review documents to learn the correlation between the source and target domains. 
CDRIB \cite{cdrib} attempts to capture the domain-shared information via information bottleneck \cite{ibp}. 
Compared with them, CDRNP has important design differences that we capture the preference correlations among overlapping and cold-start users and utilize explicit user-item interactions from both domains.

\textbf{Meta-learning} is a learning-to-learn paradigm, aiming to improve the performance of new tasks through training on a distribution of related tasks. Inspired by this, meta-learning methods have been introduced to alleviate the cold-start problem in recommender systems, which include
optimization-based methods~\cite{mamo,melu,mlhin}, gradient-based methods \cite{mamlfa,reptile,mlmann} and model-based methods~\cite{cnp,np,oslhnb}.
MeLU~\cite{melu} fit a candidate selection strategy into a standard model-agnostic meta-learning framework to determine distinct items for customized preference estimation. MAML~\cite{mamlfa} aims at learning model’s initial parameters which are capable of adapting to new tasks with only a few gradient steps. The proposed CDRNP falls into the model-based methods, which utilizes NP~\cite{np} to capture the preference correlations among users.

\textbf{Neural Process} is a well-known member of stochastic processes family. On the one hand, NP can model the distribution over functions like Gaussian processes. On the other hand, NP is structured as neural networks which can generate predictions in a efficient way. NP are originally designed to the tasks of few-shot image completion~\cite{cnp,np}, image recognition~\cite{npmatch}.
Since the first model CNP~\cite{cnp}, there have been numerous subsequent works to improve NP in various aspects. ANP~\cite{anp} introduces a self-attention mechanism to improve the midel's reconstruction quality. BNP~\cite{bnp} uses bootstrap to learn the stochasticity in NP without assuming a particular form. Recently, a series of methods~\cite{tanp,idnp,lieinp} utilizing NP have emerged in the field of recommender systems. 
TANP utilizes Neural Process to solve the cold-start issue. However, TANP focuses on single-domain recommendation, which has quite different task setting comparing to cross-domain recommendation.

\section{Conclusion}
In this paper, we propose a novel meta-learning-based method CDRNP for CDR to cold-start users via Neural Process. 
CDRNP captures the preference correlations among overlapping and cold-start users with an associated stochastic process.
Additionally, CDRNP introduces a preference remainer for enhancing the common preference from the overlapping users, and an adaptive conditional decoder to make prediction. 
Experimental results demonstrate that CDRNP consistently outperforms SOTA methods in three CDR scenarios. 
In the future, we will further investigate Neural Process for multi-domain recommendation.

\section*{Acknowledgement}
This work was funded by the National Key Research and Development Program of China (No.2021YFB3100600), the Strategic Priority Research Program of Chinese Academy of Sciences (No.XDC02040400), the Youth Innovation Promotion Association of CAS (No.2021153), and the National Postdoctoral Program (No.GZC20232968).

\balance
\bibliographystyle{ACM-Reference-Format}
\bibliography{sample-base-extend.bib}
\end{document}